\documentclass[aps,pre,twocolumn,showpacs]{revtex4}
\usepackage{amsmath,amsthm}  
\usepackage{epsf}
\usepackage{amssymb}  
\usepackage{graphicx}

\newcommand{\be}{\begin{eqnarray}}
\newcommand{\ee}{\end{eqnarray}}
\newcommand{\eq}[1]{Eq.~(\ref{#1})}
   
\newcommand{\twomatrix}[4]{\left( \begin{array}{cc} #1 & #2 \\ #3 & #4 \end{array}\right)}
\begin{document}
\title{Excited-State Quantum Phase Transitions in Dicke Superradiance Models}
\author{Tobias Brandes}
\affiliation{Institut f\"ur Theoretische Physik,
  Hardenbergstr. 36,
  TU Berlin,
  D-10623 Berlin,
  Germany}
\date{\today{ }}
\begin{abstract}
We derive analytical results for various quantities related to the excited-state quantum phase transitions in a class of 
Dicke superradiance models in the semiclassical limit. Based on a calculation of a partition sum restricted to Dicke states, we discuss the singular behavior of  the derivative  of the density of states and find observables like the mean (atomic) inversion and the boson (photon) number and its fluctuations at arbitrary energies.
Criticality depends on energy and a parameter that quantifies the relative weight of rotating versus counter-rotating terms, and 
we find a close analogy to the logarithmic and jump-type non-analyticities known from the Lipkin-Meshkov-Glick model. 
\end{abstract}

\maketitle

\section{Introduction}
The recent successful experimental realization \cite{Dicke_experiment} 
of the Dicke-Hepp-Lieb  superradiance \cite{Dic54,HL73} with cold atoms in photonic cavities has  sparked a renewed interest in the Dicke model. Although a detailed understanding of the quantum  phase transition (QPT) associated with the phenomenon requires somewhat 
more involved modeling \cite{PhysRevA.75.013804,NKSD10,PiazzaStrackZwerger13,BMSK2012,RDBE13}, the simplest one-mode form of the Dicke Hamiltonian (a boson coupled to a large angular momentum) continues to serve as a simple model with fascinating properties. One reason for this is the non-integrability of the model 
and the appearance of quantum chaos and its relation to the bifurcation-type QPT in the thermodynamic limit $N\to \infty$
of infinitely many (pseudo-spin $\frac{1}{2}$) two-level systems  \cite{EB03two,Bra05,AH12,AH_NJP12}.

Apart from various modifications of the  model for adaptation to, e.g., multi-level systems \cite{Hayn,BNC13} or realizations in other materials \cite{Nataf_Ciuti,CPGM12,DeLiberato}, the Dicke model has been discussed recently \cite{PeresFernandezetal2011,PeresFernandezetal2011Jz,PRR13} in the context of excited-state quantum phase transitions (ESQPT). In contrast to ground state QPTs, these occur at higher energies and have singularities in the energy level structure as their hallmark \cite{Cejetal2006,CCI08}. The numerical calculations  \cite{PeresFernandezetal2011} confirmed a line of ESQPTs in the superradiant phase of  the Dicke model at the energy coinciding with the ground state energy of the  normal phase. In a semi-classical picture, this energy  corresponds to the excitation energy from a spontaneously symmetry-broken ground state right onto the top of a local maximum in a Landau functional-type potential. 

Most of the research on ESQPT so far has been dealing with mean-field type Hamiltonians, where  a classical potential landscape governs the singularities for both types of quantum phase transitions (cf. \cite{PeresFernandezetal2011} for further references).
This is also the case for the Lipkin-Meshkov-Glick (LMG) model \cite{PeresFernandezetal2009} that describes a simple non-linearity for a large angular momentum and for which Ribeiro, Vidal, and Mosseri \cite{RVM07,RVM08} presented an exhaustive analysis of the phase diagram. Our results for the class of Dicke superradiance models discussed in this paper reveal a very close analogy to their findings for the LMG model, but they also show interesting aspects that are particular to the superradiance case and highlight the role of the semiclassical limit $N\to \infty$, $\hbar\to 0$ at constant $N\hbar$ \cite{AH12,AH_NJP12} for the level density $\nu(E)$ and all quantities derived from it.

In our model we introduce a control parameter $g$ that quantifies the relative weight of rotating versus counter-rotating terms \cite{Hioe73,ALS07} which corresponds to the amount of anisotropy in the LMG model coupling parameters.
As a limit of particular interest we then obtain the Tavis-Cummings model \cite{Tavis}, where the integrability leads to a Goldstone mode that persists throughout the superradiant phase and that removes a logarithmic ESQPT singularity in favor of a `first order' jump-type discontinuity, with a re-emerging of the former if the Hilbert space is properly restricted to a single excitation number only. 

The key difference between ESQPT and finite-temperature phase transitions in quantum systems is the role of entropy. The ESQPT in the Dicke model appears in the Hilbert space spanned by the Dicke states $|jm\rangle$ with fixed $j=N/2$. The level density (density of states, DOS) $\nu(E)$ of the eigenstates of the Hamiltonian of energy $E$ then defines a micro-canonical ensemble that is abnormal in the thermodynamical sense that the associated entropy $S(E)\equiv \ln \nu(E)$ does not scale linearly with the particle number $N$, but only logarithmically, i.e. $\nu(E)$ itself is only proportional to $N$.  In contrast, the canonical partition sum $Z(T)$ that determines the original calculations \cite{WH73,HL73a} for finite-temperature phase-diagram sums over all $2^N$ states of the two-level systems, the saddle point approximation to $Z(T)$ for $N\to \infty$ contains the typical entropic contribution reflecting the high degree of degeneracy in that case, and thermodynamic quantities like free energy and entropy are extensive, i.e. proportional to $N$. 

The structure of this paper is a follows: section II describes the model and the main method, section III discusses some general properties of the level density $\nu(E)$, and section IV is devoted to a detailed discussion of the Dicke model. In Section V we then describe the ESQPTs in the generalized Dicke models, in section VI the somewhat exceptional case of the restricted Tavis-Cummings model, and we conclude in section VII. The appendices A-D give some technical details on the angular momentum traces, the integrations needed in the Dicke model, the logarithmic singularities, and the Bogoliubov transformation for the normal phase of the generalized Dicke models.

\section{Model and Method}

\subsection{Hamiltonian}
Our model describes a single bosonic cavity mode $a^\dagger$ coupled to $N$ two-level systems that are described by a 
collective angular momentum algebra $J_\alpha\equiv \frac{1}{2}\sum_{j=1}^N\hat{\sigma}_{\alpha}^{(j)}$, $\alpha=x,y,z$ with 
$J_\pm\equiv  J_x \pm i J_y$ 
and Pauli-matrices $\hat{\sigma}_{\alpha}^{(j)}$.
The Hamiltonian reads
\be\label{Hamiltonian}
\mathcal{H}=\hbar\omega a^\dagger a + \hbar \omega_0J_z+ \frac{\hbar\lambda}{\sqrt{N}}\sum_{\pm} \frac{1\pm g}{2}\left ( a J_\pm + a^\dagger J_ \mp \right),
\ee
where $0\le g \le 1$ is a parameter weighting rotating and counter-rotating terms such that
for $g=0$,  $\mathcal{H}$ describes the non-integrable Dicke model (rotating and counter-rotating terms) whereas $g=1$ describes
the integrable Tavis-Cummings model (rotating terms only). 
The model has a ground state QPT when the criticality parameter
\be\label{mudef}
\mu\equiv \frac{\lambda^2}{\omega \omega_0}
\ee
equals unity, with the transition from the normal phase ($\mu<1$) to the  superradiant phase $\mu>1$ . Importantly, for our 
choice of coupling \cite{footnote_coupling}
 the ground state QPT and all ground state quantities are  independent of the value of $g$ \cite{Hioe73}, whereas $g$ will turn out as a control parameter for the
ESQPT in the superradiant phase. 

In the particular case  $g=1$ (Tavis-Cummings model), the Hamiltonian $\mathcal{H}$ conserves the excitation number
\be\label{Nexdef}
N_{\rm ex} \equiv a^\dagger a + J_z + j,
\ee
where again $j=N/2$ and one has to specify the value(s) of $N_{\rm ex}$ for which the QPT is discussed. In contrast, the $g\ne 1$ case only conserves a parity defined by $(-1)^{N_{\rm ex}}$.

We will include the limit $g=1$ in the  discussion for $0\le g<1$ below in a natural way by defining an `unrestricted' Tavis-Cummings model where the calculation is performed by averaging over all values of $N_{\rm ex}$. In the last section, we then come back to the Tavis-Cummings model restricted by a fixed excitation number, which turns out to be technically somewhat more involved than the unrestricted case. If fact, QPT criticality is (for our choice of coupling strengths) determined by the condition $\lambda >|\omega_0-\omega|/2$ in that case regardless of $N_{\rm ex}$ \cite{PeresFernandezetal2011}.

\subsection{Partition sum}
The key quantity to  obtain the level density (density of states, DOS) $\nu(E)$ is the partition sum $\mathcal{Z}(\beta)$, 
\be\label{ZLaplace}
\mathcal{Z}(\beta)\equiv{\rm Tr} e^{-\beta\mathcal{H}}\equiv \int_{E_0}^\infty dE e^{-\beta E} \nu(E),
\ee
from which $\nu(E)$ follows via Laplace back-transformation. Here, $E_0$ is the ground state energy of $\mathcal{H}$ \cite{footnote_Laplace}.

We evaluate  $\mathcal{Z}(\beta)$  by the method of Wang and Hioe \cite{WH73} using coherent photon states and the limit $N\to \infty$ (which we will always consider in the following),
whereby the operators $a$, $a^\dagger$ can be replaced by numbers $\alpha$, $\alpha^*$ and one obtains 
\be\label{partition}
 \mathcal{Z}(\beta) &=& \int \frac{d^2\alpha}{\pi}e^{-\beta \hbar \omega |\alpha|^2} 
Z(\alpha;\beta)\nonumber\\
Z(\alpha;\beta)&\equiv&{{\rm Tr}}
 e^{-\beta \hbar \left[ \omega_0J_z+ \frac{\lambda}{\sqrt{N}}\sum_{\pm} \frac{1\pm g}{2}\left ( \alpha J_\pm + \alpha^* J_ \mp \right)\right]}.
\ee
The next step is to evaluate the angular momentum trace $Z(\alpha;\beta)$ (which is taken over the basis of Dicke states $|jm\rangle$ with maximum $j=N/2$ only \cite{footnote_Dicke_states}) by employing the semiclassical limit
$\beta \hbar \omega_0 \to 0$. In the energy domain, this means that we are interested in energies $E$ that are macroscopic with respect to  $\hbar \omega_0$ in the sense that the scaled energy
\be\label{varepsilondef}
\varepsilon \equiv \frac{E}{N \hbar \omega_0}
\ee
is of order one. The semiclassical limit is thus formally defined as $\hbar\to 0$ together with the thermodynamic limit $N\to \infty$ such that the product $L\equiv \hbar j=\hbar N/2$ remains finite, where $L$ is the conserved classical angular momentum \cite{AH_NJP12}.

The evaluation of $Z(\alpha;\beta)$ (Appendix A) yields
\be\label{partition2}
& &\mathcal{Z}(\beta)=  
\int_1^\infty dy \int_0^{2\pi}d\varphi
\frac{N
 \sum_\pm \pm 
e^{ N \beta \hbar \omega_0 \Phi_\pm\left( \alpha(\varphi),y\right)  }}{ 4\pi \mu \beta \hbar \omega g},
\ee
where we defined the two dimensionless functions
\be\label{Phidef}
\Phi_\pm(\alpha,y) \equiv -\frac{\alpha}{4}(y^2-1) \pm \frac{y}{2}
\ee
that play an important role in the following analysis, 
and the $\varphi$-dependent  function
\be
\alpha(\varphi) \equiv \frac{1}{\mu} \left( 1 + \frac{1-g^2}{g^2}\sin^2 \varphi\right).
\ee 

\subsection{Density of states}
The density of states $\nu(E)$ is determined  by the inverse Laplace transformation
\be
\mathcal{L}^{-1}\left[\frac{e^{ N \beta \hbar \omega_0 \Phi_\pm}}{\beta}\right](E) =\theta(E+N \hbar \omega_0 \Phi_\pm )
\ee
under the integrals in  \eq{partition2}, from which we obtain our first key expression  
\be\label{nukey1}
\nu(\varepsilon) &=& 
\frac{N}{\hbar \omega}   \int_0^{2\pi} \frac{d\varphi}{4\pi \mu g}
I(\varphi)\nonumber\\
 I (\varphi)&\equiv& \sum_\pm \pm\int_1^\infty dy \theta \left( \varepsilon
+\Phi_\pm(\alpha(\varphi),y)\right),
\ee
with the unit-step function $\theta(x)$. We thus find $\nu(\varepsilon)$ as a product of the `system size' $N=2j$, the 
constant DOS  $1/\hbar\omega$ (corresponding to the semiclassical limit $\beta\hbar\omega\to 0$ of a single oscillator mode
with frequency $\omega$), and a term that only depends on the dimensionless  energy $\varepsilon$, \eq{varepsilondef}, and the parameters $\mu$, \eq{mudef}, and $g$.

\section{Limiting cases of $\nu(\varepsilon)$}

The physics contained in the level density $\nu(\varepsilon)$ is quite rich, and it is therefore 
instructive to consider limiting cases before analysing the full analytical expressions to be derived from \eq{nukey1}.

\subsection{Noninteracting case}
Already the noninteracting case $\lambda=0$ shows some general features  of $\nu(E)$ that persist in the interacting case, too. We directly obtain the DOS by the inverse Laplace transformation 
of \eq{partition}, or via $\nu(E)=\sum_{n=0}^\infty \sum_m \delta(E-\hbar\omega n -\hbar\omega_0 m)$ again using the limits
$N\to \infty$ and $\hbar \to 0$, thus converting sums into integrals,
\be\label{nunonint}
\nu(\varepsilon) = \frac{N  }{\hbar \omega } \left\{  \begin{array}{cc} 0, &  \varepsilon\le -\frac{1}{2}   \\    
 \frac{1}{2}+  \varepsilon, &  |\varepsilon|<\frac{1}{2} \\
 1 &   \varepsilon\ge \frac{1}{2}
\end{array} \right.,
\ee
where $\varepsilon = -\frac{1}{2} $ is the scaled ground state energy for $\lambda=0$ (no bosons, all lower atomic levels occupied). Above this lower band-edge, $\nu(E)$ grows linearly with a slope $(\hbar^2\omega\omega_0)^{-1}$, followed by a constant DOS $N/\hbar\omega$ when $E>N\hbar\omega_0/2$, the total energy  of the upper atomic levels. Graphically, $\nu(\varepsilon)$
is very close to the weak coupling ($\mu=0.2$) curve in Fig. \ref{FignuDicke}. 

The simple form \eq{nunonint} of $\nu(\varepsilon)$ also follows from the convolution $\nu(E)=\int_{-\infty}^\infty dE' \nu_{\rm osc}(E-E') \nu_{\rm ang}(E')$ with the boson and angular momentum DOS, $\nu_{\rm bos}(E) = \theta(E)/\hbar\omega$ and  $\nu_{\rm ang}(E) = \theta(\hbar\omega_0/2 - |E/N|)/\hbar\omega_0$ in the semiclassical limit. 

\subsection{Band edges}
As a matter of fact, even for arbitrary $g$ and $\mu$ one has  
\be \label{denuzero}
\nu(\varepsilon)=\frac{N}{\hbar\omega},\quad \varepsilon\ge \frac{1}{2}.
\ee
To prove \eq{denuzero}, we note that the argument of the step-function in \eq{nukey1} is a downwards parabola as a function of $y$ with zeroes
\be\label{yzeroes}
y_{1,2}^\sigma &\equiv& \frac{\sigma}{\alpha(\varphi) } \mp \sqrt{\frac{1}{\alpha(\varphi) ^2} +1 + \frac{4\varepsilon}{\alpha(\varphi) }},\quad \sigma=\pm,
\ee
where the index $1$ ($2$) belongs to the negative (positive) root. 
For energies  below   $\varepsilon = \frac{1}{2}$,
we have $y_1^- < y_2^- <1$, and only the plus-part in the sum $\sum_\pm$ contributes to $\nu(\varepsilon)$ in \eq{nukey1}. On the other hand, 
for energies   above   $\varepsilon = \frac{1}{2}$, we find
$y_1^+<-1$, $y_2^+>1$, $y_1^-<-1$, $y_2^->1$ and the $y$-integral is given by
$
I (\varphi) = \sum_\pm \pm (y_2^\pm -1) = {2}/{\alpha}(\varphi)$.
For the remaining $\varphi$-integration we now use 
\be\label{varphiintegral}
\int_0^\pi d\varphi \frac{1}{a -b\sin^2\varphi}=\frac{\pi}{\sqrt{a}\sqrt{a-b}}
\ee
to find \eq{denuzero}
at arbitrary $g$ and $\mu$.  

The energy $\varepsilon= \frac{1}{2}$ thus plays the role of an upper edge for non-trivial behavior of the DOS $\nu(\varepsilon)$.  For $\varepsilon>1/2$,  $\nu'(\varepsilon)$ vanishes and the DOS is solely determined by the oscillator frequency $\omega$. In particular, in this high-energy limit $\nu(\varepsilon)$  does not depend on the  two-level energy $\hbar\omega_0$. 
Note that the non-analyticity of $\nu(\varepsilon)$  at $\varepsilon= \frac{1}{2}$ resulting from \eq{denuzero} is not an ESQPT, as it is not related to the interaction between the boson and the 
two-level systems: it also occurs for $\lambda=0$, where it reflects the disparity between the unbounded boson and the bounded angular momentum DOS.  

The lower band edge, on the other hand, is determined by the ground state energy, which is given by $E=-N\hbar\omega_0/2$ ($\varepsilon=-\frac{1}{2}$) in the normal phase, and 
\be\label{lowerbandedge}
\varepsilon_0\equiv \frac{E_0}{N\hbar\omega_0} \equiv -\frac{1}{4}\left( \mu + \frac{1}{\mu}\right)
\ee 
in the superradiant phase. This (known) result for $E_0$ also follows from the asymptotic behavior of the partition sum $\mathcal{Z}(\beta)$ for $\beta \to \infty$. 

The interval $[\varepsilon_0,\frac{1}{2}]$,  which we will call the `band' for the rest of the paper, therefore defines the region of non-trivial excited-state physics for the $N\to \infty$ limit of our model.

\subsection{Unrestricted Tavis-Cummings model}
For $g=1$, the $\varphi$-integration in \eq{nukey1} is trivial and the result for $\nu(\varepsilon)$ is determined by the boundaries of the $y$-integral due to the step-function, which can be expressed in terms of the zeroes \eq{yzeroes}. As a result, we find $\nu(\varepsilon)= N(y_2^+-y_1^+)/(2 \hbar\omega \mu) $ for $\varepsilon_0\le \varepsilon\le -1/2$, 
$\nu(\varepsilon)= N(y_2^+-1)/(2  \hbar\omega \mu) $ for $ -1/2\le \varepsilon\le 1/2$, and thus inside the band
\be\label{nuTavis}
\nu(\varepsilon) &=&  \frac{2N}{\sqrt{\mu} \hbar\omega}
\sqrt{\varepsilon-\varepsilon_0} \theta(\mu-\mu_c),\quad  
\varepsilon_0\le \varepsilon<-\frac{1}{2}\\
 &=&\frac{N}{2 \hbar\omega} \left[ \left(1-\frac{1}{\mu}\right) + \frac{2}{\sqrt{\mu}}
\sqrt{\varepsilon-\varepsilon_0}\right], \quad  |\varepsilon|<\frac{1}{2}.\nonumber
\ee
The DOS \eq{nuTavis} in the superradiant case is shown in Fig. \ref{FignuDicke}. Two particular features (discussed in detail in section \ref{section_g}) are clearly visible already: first, there is a jump of the derivative at $\varepsilon=-\frac{1}{2}$ which is a signal of a first order ESQPT with jump-discontinuity. Second, the infinite slope $\nu'(\varepsilon_0)$ at the lower band edge is due to the vanishing of one of the collective excitation modes above the ground state in the superradiant phase.

\subsection{Dicke model in ultrastrong coupling limit}

The limit $\mu\gg 1$ (or alternatively $\omega_0\to 0$) for the Dicke model ($g=0$) can be extracted from the exact results (see below), but also in a much simpler way via the polaron transformed Hamiltonian \cite{ABEB12} with a factorizing partition sum;
\be\label{ultra_partition}
\mathcal{Z}(\beta) = \sum_{m=-N/2}^{N/2}  e^{\frac{\beta \hbar \lambda^2}{N\omega}m^2  } \sum_{n=0}^\infty e^{-\beta \hbar \omega n}.
\ee
In the semiclassical limit $N\to \infty$, $\hbar\to 0$ we convert the sums into integrals and the DOS becomes 
\be
\nu(E)= 
\frac{N}{\hbar \omega} \int_{-\frac{1}{2}}^{\frac{1}{2}} dx
\theta \left( \frac{E}{N} + \frac{\hbar\lambda^2}{\omega}x^2\right),
\ee
which leads to the simple square-root form
\be\label{nuultrastrong}
\nu(E) =N\left( \frac{1}{\hbar\omega} - \frac{2}{\hbar\lambda} \sqrt{\frac{-E}{N\hbar\omega}}\right),\quad E_0\le E\le 0.
\ee
In this limit, the ground state energy becomes $E_0= - N\hbar\lambda^2/(4\omega)$ and the upper band edge energy $E=0$. 
\section{Dicke Model ($g=0$)}\label{section_Dicke}

\begin{figure}[t]
\includegraphics[width=\columnwidth]{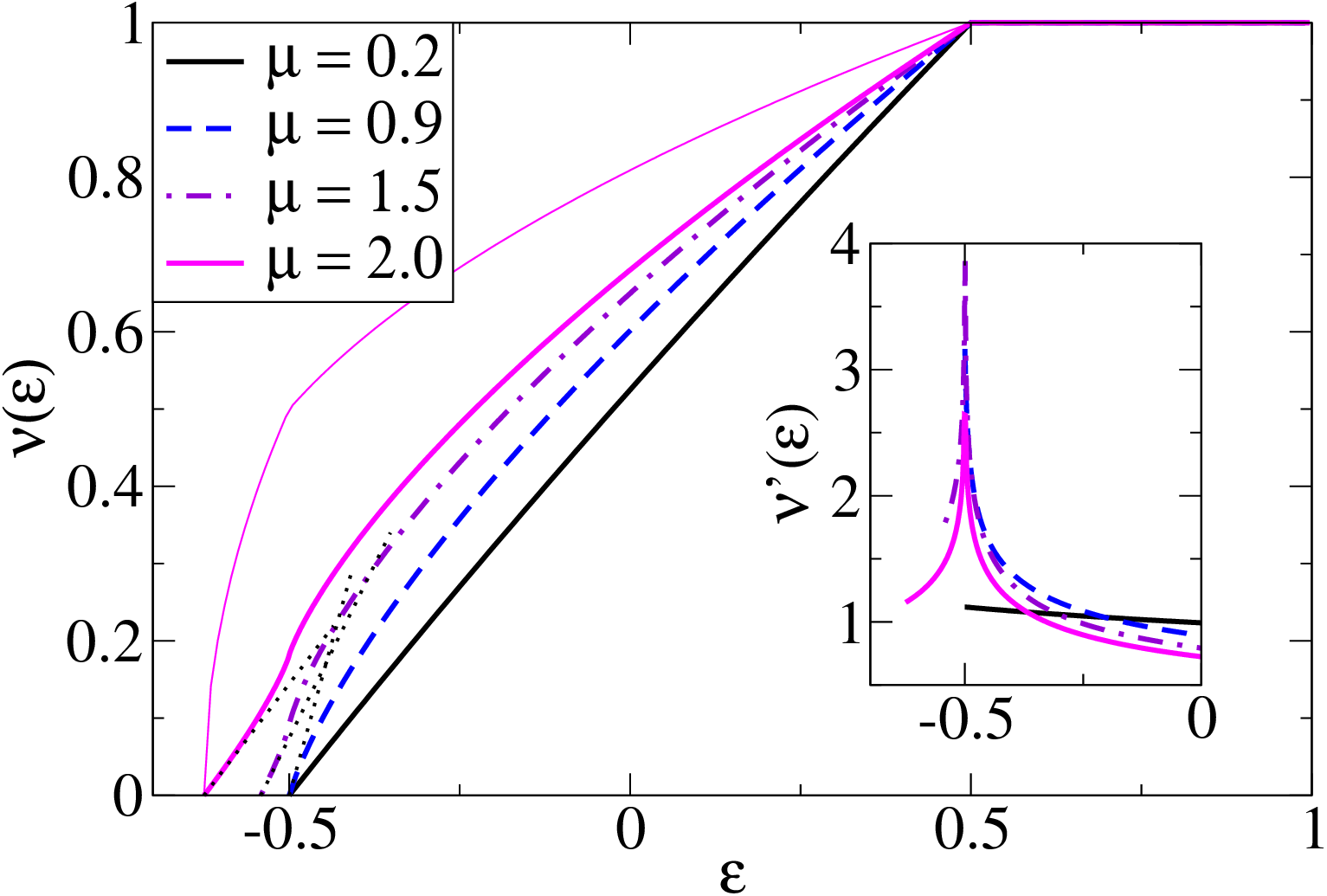}
\caption[]{\label{FignuDicke}. DOS $\nu(\epsilon)$ (in units $N/(\hbar\omega)$) as a function of scaled energy $\varepsilon\equiv E/(N\hbar\omega_0)$ 
for the Dicke model ($g=0$) at different values of  criticality parameters $\mu\equiv \lambda^2/(\omega\omega_0)$. Thin solid line: DOS $\nu(\epsilon)$ for unrestricted  Tavis-Cummings
model, \eq{nuTavis}, at $\mu=2$. Dotted lines indicate the slopes that are determined by the product of the collective excitation energies $\epsilon_\pm$
near the lower band-edge, \eq{nulowDicke}.
Inset: 
DOS derivative $\nu'(\varepsilon)$ (in units $N/(\hbar^2\omega\omega_0)$ ) displaying the logarithmic ESQPT singularity at $\varepsilon=-\frac{1}{2}$ for $\mu>1$.}
\end{figure} 
In the following, we derive and discuss results for the Dicke case $g=0$ separately because of its high relevance for the 
existing theoretical and experimental literature.

The partition sum is obtained along the lines of the calculation in section II and follows as
\be\label{partitionDicke}
& &\mathcal{Z}(\beta)=\sqrt{\frac{N}{ \pi  \beta^3\hbar^3 \omega \lambda^2}}
 \int_1^\infty dy \frac{\sum_\pm \pm e^{ N \beta \hbar\omega_0\Phi_\pm\left(\frac{1}{\mu},y\right) }}{\sqrt{y^2-1}}.
\ee
\subsection{Low-energy excitations}
We make a first interesting observation in a simple analysis of $\mathcal{Z}(\beta)$ with the Laplace method \cite{Bender_Orszag} for $\beta N\to \infty$,  
\be
\mathcal{Z}(\beta)&\sim& \frac{e^{\beta \hbar\frac{N\omega_0}{2} }}{ \beta^2 \hbar^2 \omega\omega_0\sqrt{1-\mu}} ,\quad \mu<1\\
&\sim&2 \frac{e^{\beta\hbar \frac{N\omega_0}{4}  \left( \mu + \frac{1}{\mu}\right)}}{ \beta^2  \hbar^2 \omega\omega_0 \sqrt{\mu^{2}-1} } ,\quad \mu>1.
\ee
The DOS corresponding to this asymptotic form generalizes the straight-line behavior 
\eq{nunonint} to the interacting case at  low energies $E$ and is given by
\be\label{nulowDicke}
\nu(\varepsilon) \approx N \frac{1+\theta(\mu-1)}{\epsilon_+\epsilon_-}(\varepsilon-\varepsilon_0) \theta(\varepsilon-\varepsilon_0),\quad \varepsilon\to \varepsilon_0,
\ee
with the  product $\epsilon_+\epsilon_-$ of the  excitation energies coinciding with those obtained,  e.g., via the Holstein- Primakov transformation \cite{EB03two} method. The additional factor $2$ in the superradiant phase reflects the two-fold
degeneracy of energy levels for $N\to \infty$ \cite{PRR13}, cf. the discussion in section \ref{section_degeneracies}. 

The form \eq{nulowDicke} confirms that the low-energy behaviour of the Dicke model is governed by two independent collective modes. The partition sum of the two oscillators describing these modes factorizes, and as a consequence the associated DOS for excitations above the ground state with energy $E_0$ (including  an additional degeneracy factor $g_{\rm d}$) is 
\be\label{lowenergy}
\nu_{\rm coll}(E) &=& g_{\rm d}\int dx_1dx_2\delta(E-x_1\epsilon_+-x_2\epsilon_--E_0)\nonumber \\
&=&g_{\rm d}\frac{E-E_0}{\epsilon_+\epsilon_-},\quad E>E_0
\ee
as in \eq{nulowDicke} with $g_{\rm d}=1$ in the normal and $g_{\rm d}=2$ in the superradiant phase.

\subsection{Density of states}
We will give explicit analytical results for the derivative of the  DOS within the band  in terms of an elliptic integral below, but for the numerical evalation 
it is more convenient to use the integral representation that follows from the simple Laplace back-transformation of the partition sum \eq{partitionDicke},
\be\label{nuDickeintegral}
\nu(\varepsilon) = \frac{N}{\pi\hbar\omega\mu}\int_{y_0}^{y_+} 
dy \frac{\sqrt{(y_- -y )(y-y_+)}}{\sqrt{y^2-1}},
\ee
with $y_\pm \equiv \mu \pm 2\sqrt{\mu}\sqrt{\varepsilon-\varepsilon_0}$ and the lower limit $y_0\equiv y_-$ if $\varepsilon<-\frac{1}{2}$ and
$y_0\equiv 1$ if $\varepsilon>-\frac{1}{2}$, cf. Appendix B.

In Fig. \ref{FignuDicke}, the transition from an almost straight-line form of $\nu(\varepsilon)$ at small couplings $\mu$ in the normal phase (resembling the non-interacting case \eq{nunonint})  to a more complex form in the superradiant phase is clearly visible. The slope $\nu(\varepsilon)$ at the  lower band edge is given by \eq{nulowDicke} and diverges at the QPT transition point $\mu=1$ due to the vanishing of one of the excitation energies there \cite{EB03two}, as expected.

The most interesting feature, however, is the signature of the  ESQPT at $\varepsilon=-\frac{1}{2}$ in the superradiant phase ($\mu>1$), a feature that was first found numerically by P\'{e}r{e}z-Fern\'{a}ndez and co-workers \cite{PeresFernandezetal2011,PeresFernandezetal2011Jz}. This non-analyticity is only weakly visible in $\nu(\varepsilon)$ itself but it shows very clearly  in the form of a logarithmic divergence in the derivative  $\nu'(\varepsilon)$ (inset). Near $\varepsilon=-\frac{1}{2}$, we find
\be\label{nuDickelog}
\nu'(\varepsilon) \approx -\frac{\log \frac{\left|  \varepsilon+\frac{1}{2} \right|}{16 (1-1/\mu)^2} }{\pi\hbar^2\omega_0\omega\sqrt{\mu-1}}
\ee
as derived in Appendix C (also cf. \eq{logdiv}). The origin of this feature lies in a saddle-point in a classical potential landscape \cite{CCI08,PeresFernandezetal2011}, cf. section \ref{potential_landscape}.

\subsection{Expectation values of observables}
\begin{figure}[t]
\centerline{\includegraphics[width=\columnwidth]{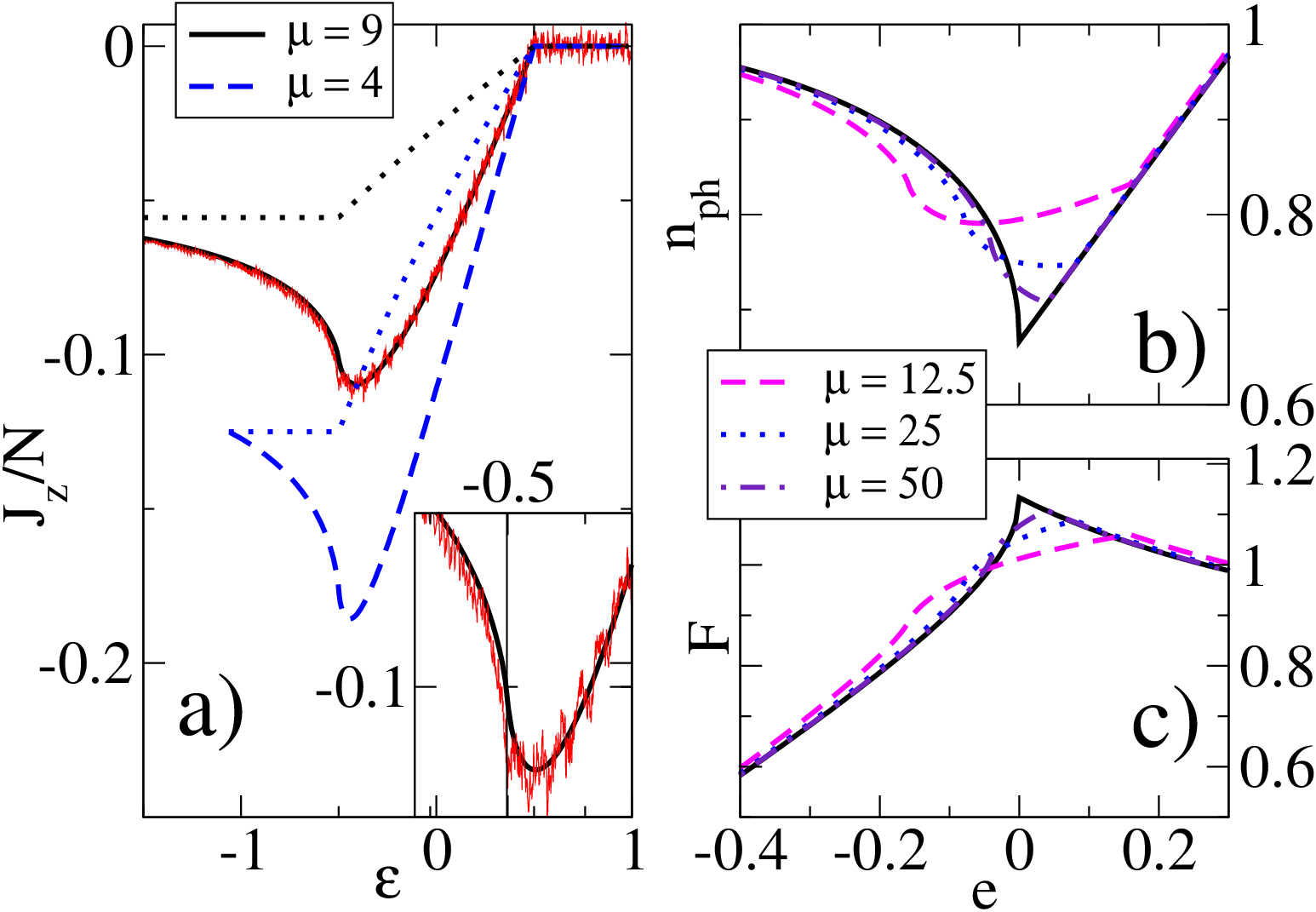}}
\caption[]{\label{Figobservables} a) Inversion $\langle J_z \rangle/N$   as a function of scaled energy $\varepsilon\equiv E/N\omega_0$ for the Dicke model $(g=0)$: analytical results from \eq{Jz} and numerical data for $\mu=9$ by P\'{e}rez-Fernand\'{e}z {\em et al.},  Fig. 2 in \cite{PeresFernandezetal2011Jz} (red fuzzy line, the inset shows a zoom near $\varepsilon=-\frac{1}{2}$). Dotted lines indicate the result \eq{JzTavisCummings} for the (unrestricted) Tavis-Cummings model ($g=1$); b) boson number $n$  scaled with its ground state values, \eq{Jzground}, and c)  Fano factor $F\equiv$ var $(n)/n$, both 
as a function of scaled energy $e\equiv E/|E_0|$   for various criticality parameters $\mu$ approaching the ultrastrong coupling regime (solid black curve), \eq{phultra}.}
\end{figure} 

As in the LMG model \cite{RVM08}, we can obtain averages of observables $\hat{A}$ such as the inversion $J_z$ or the boson number  $\hat{n}\equiv a^\dagger a$ as sums over
eigenstates $|\alpha\rangle$ with energies $E_\alpha$,
\be\label{observable1}
 \langle \hat{A} \rangle(E) = \frac{1}{\nu(E)}\sum_\alpha \langle \alpha| \hat{A}| \alpha \rangle \delta(E-E_\alpha).
\ee
This can be re-written by use of the Hellmann-Feynman theorem, which takes advantage of a parametric dependence of the $E_\alpha$, i.e., 
$\frac{\partial E_\alpha(\lambda)}{\partial \lambda}= \langle \alpha|\frac{\partial}{\partial \lambda} \mathcal{H}  | \alpha \rangle
$.  In our microcanonical ensemble, we can now re-write \eq{observable1} as 
\be\label{Jz}
\langle J_z \rangle(E) &=& \frac{1}{\nu(E)} \sum_\alpha \frac{\partial E_\alpha(\omega_0)}{\partial \hbar\omega_0}  \delta(E-E_\alpha)\nonumber\\
&=& -\frac{1}{\nu(E)}  \frac{\partial }{\partial \hbar\omega_0} \int_{-\infty}^E dE'\nu(E'),
\ee
and correspondingly for $\langle \hat{n} \rangle(E) $ with the derivative with respect to $\omega$ instead of $\omega_0$. Note that these expressions hold for all our models (arbitrary $g$). 

In the superradiant phase of the Dicke model ($g=0$), this generalizes the ground state expectation values \cite{EB03two}
\be\label{Jzground}
\langle J_z \rangle (E_0) &=&  -\frac{N}{2 \mu},\quad
\langle \hat{n} \rangle (E_0) = \frac{N}{4}\frac{\omega_0}{\omega}\left(\mu - \frac{1}{\mu}\right)
\ee
to higher energies, with \eq{Jzground} following from the  l'H\^{o}pital rule applied to \eq{Jz},
\be\label{Jzthermo}
\langle J_z \rangle (E_0)  = - \left.\frac{\frac{\partial }{\partial \hbar\omega_0}\nu(E)}{\frac{\partial }{\partial E} \nu(E)}\right|_{E=E_0} = \frac{\partial E_0}{\partial \hbar\omega_0},
\ee
and again correspondingly for $\langle  \hat{n} \rangle(E_0) $, with $E_0$ given in \eq{lowerbandedge} 
\cite{footnote_thermo}.
Explicit expressions for the integrals needed in \eq{Jz} are given in Appendix B.

Results for the inversion  as a function of energy in the superradiant regime $\mu>1$  are shown in Fig. \ref{Figobservables} a). 
First,  $\langle J_z\rangle(\varepsilon)$ becomes flat and levels off at exactly zero above the upper band edge $\varepsilon=\frac{1}{2}$, a feature that has already been present in the numerical data of P\'{e}rez-Fern\'{a}ndez {\em et al.} \cite{PeresFernandezetal2011Jz}. In fact, we have for arbitrary $g$ and $\mu$ that
\be\label{Jzzero}
\langle J_z\rangle = 0,\quad \varepsilon \ge \frac{1}{2},
\ee
which follows from \eq{Jz} where we can extend the upper limit of the integral to $\infty$ for $ E\ge \frac{N\hbar \omega_0}{2}$ because of \eq{denuzero}, and we use $\int_{-\infty}^\infty dE \nu(E) = \mathcal{Z}(\beta=0)$, \eq{ZLaplace}, with the partition sum being infinite in the  limit of infinite temperature but formally independent of $\omega_0$, cf. \eq{partition}. 

The agreement between our  analytical result and the numerical data for $j=30$ (fuzzy red line in Fig. \ref{Figobservables} for $\mu=9$ \cite{PeresFernandezetal2011Jz,Jzfit}) is so good that the curves  basically  lie on top of each other for all energies. The data from the numerical diagonalization  are obtained as an average over 20 eigenstates, but they still show quantum oscillations due to the finiteness of
$N$ (and $\hbar$). In contrast, the analytical result is based on the semiclassical limit $N\to \infty$, $\hbar\to 0$ (with $N\hbar$ kept constant) which smears out these oscillations. 

Fig. \ref{Figobservables} also shows how hard it is to directly see the logarithmic singularity at $\varepsilon=-\frac{1}{2}$ in the observable $\langle J_z\rangle$: on the scale shown in the figure, it is somewhat masked by the minimum that lies slightly  above $\varepsilon =-\frac{1}{2}$. From our analytical expressions, we extract the derivative
\be\label{Jzlog}
\frac{\partial}{\partial \varepsilon} \langle J_z\rangle(\varepsilon) \propto \log \left| \varepsilon + \frac{1}{2}\right|,
\ee
cf. Appendix B. 

The logarithmic singularity is absent in the restricted Tavis-Cummings model, where instead of \eq{Jzlog} we find  that $\langle J_z\rangle(\varepsilon) = -N/(2\mu)$ is constant  below the critical energy $\varepsilon= -\frac{1}{2}$, followed by a square-root non-analyticity, 
\be\label{JzTavisCummings}
\frac{\langle J_z\rangle(\varepsilon)}{N} &=& \frac{\varepsilon -\frac{1}{2}\left( \sqrt{\frac{4\varepsilon}{\mu}  + \frac{1}{\mu^2} + 1 } -\frac{1}{\mu}\right)}{\mu-{1}+ \mu\sqrt{\frac{4\varepsilon}{\mu}  + \frac{1}{\mu^2} + 1 } },
|\varepsilon|\le \frac{1}{2},
\ee
and  a vanishing of $\langle J_z\rangle(\varepsilon)$ above the upper band-edge $\varepsilon=\frac{1}{2}$ (dotted lines in Fig \ref{Figobservables}). The  non-analyticity at the ESQPT position $\varepsilon= -\frac{1}{2}$ is consistent with the jump in  the DOS derivative $\nu'(\varepsilon)$, cf. \eq{nuTavis} and the discussion in the next section.

\subsection{Average boson number and its fluctuations}
We can directly relate expectation values of the boson number $\hat{n}\equiv a^\dagger a$ and powers thereof to the $Q$ or Husimi function $Q(\alpha;\beta)\equiv e^{-\beta\hbar \omega |\alpha|^2}Z(\alpha;\beta) $ that appears as integrand in our partition sum $\mathcal{Z}(\beta)$, \eq{partition}. In the semiclassical limit, normal or anti-normal ordering of operators plays no role and we can write
the definition \eq{observable1} as
\be\label{QLaplace}
\langle \hat{n}^m \rangle(E) = \frac{1}{\nu(E)}  \mathcal{L}^{-1}\left[\int \frac{d^2\alpha}{\pi} |\alpha|^{2m}Q(\alpha;\beta)\right](E)
\ee
with the inverse Laplace transform of the $Q$ function. The $|\alpha|^{2m}$ under the integral can be replaced by the operation $(-1)^m\frac{\partial^m}{ \partial (\beta\hbar\omega)^m}$ which is useful to derive explicit results. For $m=1$, we thus immediately recover the Hellmann-Feynman form \eq{Jz} for $\langle \hat{n}\rangle(E)$ (division of $ \mathcal{Z}(\beta)$ by $\beta$ corresponds to integration over energy of $\nu(E)$). 
Details for the integrals needed for $\langle \hat{n}^m \rangle(E)$, $m=1,2$ are given in Appendix B. 

The agreement between our analytical result for $\langle \hat{n} \rangle(E)$ and  numerical data of P\'{e}rez-Fern\'{a}ndez for the case $\mu=9$ (unpublished data for $j=15$, not shown here) is very good for all energies (numerically this requires large boson numbers for the truncated boson Hilbert space). Above the upper band edge $\varepsilon \ge \frac{1}{2}$, we reproduce the linear dependence in energy
found by Altland and Haake  via their classical $Q$-function \cite{AH_NJP12}: from  \eq{denuzero} and the limiting value at $\varepsilon=\frac{1}{2}$ we find
 $\langle \hat{n}\rangle (\varepsilon) =  N   \frac{\omega_0}{\omega} \left( \frac{\mu}{6}+ \varepsilon\right)$. Similarly, our expressions for the variance var $\hat{n}(E)\equiv(\langle \hat{n}^2\rangle  -\langle \hat{n}\rangle^2)(E)$ exactly reproduce the linear energy dependence for $\varepsilon \ge \frac{1}{2}$ \cite{AH_NJP12} and display the macroscopic scaling with $N^2$ at any finite energy $E>E_0$. At $E=E_0$, the variance vanishes as our calculation only accounts for the leading terms  $\propto N^2$ and is not sensitive to subleading dependencies $\propto N$ \cite{Casetal11}.
As expected, the ESQPT log-singularity shows up at $\varepsilon=-\frac{1}{2}$ in the mean value and the variance.

In the ultrastrong Dicke limit \eq{nuultrastrong}, from the partition sum \eq{ultra_partition} and \eq{QLaplace} we derive the  mean value and the variance of the scaled boson number $\hat{n}/\langle a^\dagger a \rangle(E_0)$,
\be\label{phultra}
n(e<0) &=& \frac{\frac{2}{3}+e + \frac{1}{3} (-e)^{\frac{3}{2}}}{1-\sqrt{-e}} \\
 \mbox{\rm var } n(e<0) &=& \frac{1}{5}\left( (1+\sqrt{-e})(6 + 4 e) + e^2\right) -n^2\nonumber\\ 
n(e>0)&=& \frac{2}{3} + e,\quad  \mbox{\rm var } n(e>0) = \frac{34}{45} + \frac{2}{3}e\nonumber ,
\ee
with the  energy variable $e\equiv E/|E_0|>-1$ scaled with the ground state energy $E_0$. Fig. \ref{Figobservables} b), c) clearly shows \eq{phultra} as the limiting form for the scaled boson number $n$
and the Fano factor $F\equiv \mbox{\rm var} n /n $ when $\mu$  increases to large values in the superradiant regime.

\section{Generalized Dicke Models ($g\ge 0$)}\label{section_g}
We now turn to the general case of arbitrary $0\le g\le 1$ in our model Hamiltonian $\mathcal{H}$, \eq{Hamiltonian}. 

\subsection{Classical Potential}\label{potential_landscape}
As we are dealing with a mean-field Hamiltonian in the thermodynamic limit $N\to \infty$, all critical features are expected to be related to extremal points in a classical potential landscape \cite{CCI08}. In fact, the following very simple analysis of potential extrema is very helpful for interpreting the various critical regions following from the exact expressions for the DOS derivative $\nu'(E)$ in terms of elliptic integrals.

 The partition sum $\mathcal{Z}(\beta)$, \eq{partition}, contains a potential in a natural way: after carrying out the angular momentum trace, we can write it as phase-space integral;
\be\label{potential}  
\mathcal{Z}(\beta) &=& \frac{N}{\pi \beta\hbar \omega} \int dx dp \frac{\sum_\pm \pm e^{-\beta N\hbar \omega_0 U_\pm (x,p)}}{\sqrt{1 + \mu (x^2+p^2)}}\nonumber\\
U_\pm(x,p) &\equiv& \frac{1}{4}(x^2+p^2) \mp \frac{1}{2}\sqrt{1 + \mu (x^2 + g^2 p^2)}.
\ee
The contribution relevant for the region $\varepsilon\le\frac{1}{2}$ below the upper band-edge of the DOS is the plus part, i.e. $U_+$, whereas the  minus part, i.e. $U_-$, only contributes 
to $\varepsilon\ge\frac{1}{2}$ in $\nu(\varepsilon)$ leading to the levelling-off at the constant oscillator DOS, cf. \eq{denuzero}.  

The extrema of the (dimensionless) potential $U_+(x,p)$ have a simple structure. We only discuss the superradiant phase $\mu\ge 1$ which is of interest for the ESQPT. For all $g$, there are two minima
at $(x=\pm \sqrt{\mu- 1/\mu},p=0)$ where
$U(x,p)=\varepsilon_0$ (the scaled ground state energy \eq{lowerbandedge} as expected) and an extremum at $(x=0,p=0)$ where $U(x,p)=-1/2$, 
the scaled ESQPT critical energy. For $g<\frac{1}{\sqrt{\mu}}$, the extremum $(x=0,p=0)$ is a saddle point leading to logarithmic non-analyticities in $\nu(\varepsilon)$ as we already saw in the Dicke case $g=0$. 

In contrast, for $g>\frac{1}{\sqrt{\mu}}$, the saddle point at $(x=0,p=0)$ is transformed into a local maximum, and instead two new saddle points at finite momenta, $(x=0,p=\pm \sqrt{g^2\mu - 1/(g^2\mu)})$ appear where 
$U(x,p)=\varepsilon_g$ with the energy $\varepsilon_g$ (again scaled by $N\hbar\omega_0$) given by
\be\label{epsgdef}
\varepsilon_g\equiv -\frac{1}{4}\left(g^2 \mu+ \frac{1}{g^2\mu}\right).
\ee
As we will see below, this leads to a log-type ESQPT in $\nu'(\varepsilon)$ at $\varepsilon=\varepsilon_g$, in addition to a non-analyticity at  $\varepsilon=-1/2$ that is now a first order, jump discontinuity type ESQPT.

Finally, at $g=1$ (restricted Tavis-Cummings model), the potential becomes a Mexican hat with only one local maximum at $(x=0,p=0)$ and a continuous ring of minima  where again $U(x,p)=\varepsilon_0$. We emphasize that 
 this Goldstone mode appears for all couplings $\mu>1$ in the superradiant phase and not only at criticality \cite{ALS07}.
Its origin lies in the gauge-symmetry $a\to ae^{i\phi}$, $J_+\to e^{-i\phi}J_+$ of the  Hamiltonian $\mathcal{H}$ which is in rotating-wave form at $g=1$. In the normal phase, $\langle a\rangle$ and $\langle J_+ \rangle$ vanish and this 
symmetry plays no big role in contrast to the superradiant phase where both expectation values become macroscopic.

As a consequence, for $g=1$, one of the collective excitation energies in the superradiant phase vanishes, as we also directly confirmed using the equation-of-motion method by Bhaseen  {\em et al.} \cite{BMSK2012}. This is the reason for the divergence of $\nu'(\varepsilon)$ at the lower band-edge $\varepsilon=\varepsilon_0$, as we already observed in \eq{nuTavis} and Fig. \ref{FignuDicke}. In contrast, for $g<1$ one has a non-diverging $\nu'(\varepsilon_0)$, cf. \eq{nulowDicke} for the Dicke case $g=0$ .

\subsection{Exact expressions for $\nu'(\varepsilon)$}
We now turn to the full exact solution for  arbitrary $0\le g\le 1$.
Instead of trying a  direct evaluation of the DOS $\nu(\varepsilon)$, \eq{nukey1} (which is cumbersome due to the step-function), 
progress is made by calculating the derivative $\nu'(\varepsilon)$, for which we obtain the  expression
\begin{widetext}
\be\label{nu1}
 \nu'(\varepsilon) &=& 
\frac{1}{4\pi \hbar^2 \lambda^2 g}\int_1^\infty dy \int_0^{2\pi} d\varphi  \sum_\pm \pm \delta\left(\varepsilon +  
\Phi_\pm\left(\alpha(\varphi),y\right) \right) \nonumber\\
&=& -\frac{1}{2\pi \hbar^2  \lambda^2 g} \sum_\pm \pm \Im \int_1^\infty \frac{dy}{\sqrt{\varepsilon+i0 + \Phi_\pm(\mu^{-1},y)} \sqrt{\varepsilon+i0 + \Phi_\pm(g^{-2} \mu^{-1},y)} }
\ee
\end{widetext}
with $\Phi_\pm$ defined in \eq{Phidef}, and where we used $-\pi \delta(x)=\Im 1/(x+i0)$ and    \eq{varphiintegral}.

To evaluate \eq{nu1}, we first recall that only the plus term in the $\sum_\pm$ contributes within the band $[\varepsilon_0,\frac{1}{2}]$ (cf. the remark after \eq{yzeroes}). Next, we use $\sqrt{x+i0}=\sqrt{x}\theta(x) + i \sqrt{-x}\theta(-x)$ to re-write \eq{nu1} within the band  as 
\be
\nu'(\varepsilon) &=& \frac{1}{2\pi  \hbar^2 \lambda^2 g}\int_1^\infty \frac{dy\theta(\varepsilon -p(y)) \theta (q(y)-\varepsilon)}{\sqrt{-(\varepsilon -p(y))  (\varepsilon -q(y)) }  },
\ee
where we defined the two parabolas $p(y)\equiv -\Phi_+(\mu^{-1},y)$, $q(y)\equiv -\Phi_+(g^{-2}\mu^{-1},y)$  between which the energy $\varepsilon$ has to lie. This determines the boundaries of the $y$- integral, expressed in terms of the zeroes of $p(y)$ and $q(y)$,
\be\label{yzdef}
y_{\pm}&\equiv& \mu \pm 2\sqrt{\mu} \sqrt{\varepsilon -\varepsilon_0}\\
z_{\pm}&\equiv& g^2 \mu \pm 2\sqrt{g^2\mu} \sqrt{\varepsilon -\varepsilon_g}.
\ee

In the superradiant phase ($\mu>1$), by considering $p(y)<\varepsilon<q(y)$  we find
two regimes depending on the value of the parameter $g$.
For $g^2\mu <1$,  for energies $\varepsilon\le -\frac{1}{2}$ the boundaries are $[y_-,y_+]$ .
In contrast, for  $g^2\mu > 1$ there are two regions: one with   boundaries
$[y_-,y_+]$  if $\varepsilon\le \varepsilon_g$, and the other for energies $\varepsilon_g\le \varepsilon \le -\frac{1}{2}$ with $y_-\le z_-\le z_+\le y_+$ and two intervals $[z_+,y_+]$ and $[y_-,z_-] $ contributing to the $y$-integral.
Furthermore, for all values of $g$ and  $\mu$ and for energies $\varepsilon\ge -\frac{1}{2}$, the boundaries are $[z_+,y_+]$ with $z_-\le y_-\le z_+\le y_+$.  
We also note that for $\varepsilon\le \varepsilon_g$, the $z_\pm$ become complex. 

This now allows us to give explicit expressions for $\nu'(\varepsilon)$ \cite{Gradstein}
\begin{widetext}
\begin{subequations}
\label{denuexact}
\begin{align}
 \varepsilon_0\le\varepsilon\le\varepsilon_g&:
\quad \nu'(\varepsilon) = 
\frac{4}{\pi \hbar^2 \omega\omega_0}
\frac{1}{\sqrt{p_+p_-}} K\left(\frac{(y_+-y_-)^2-(p_+-p_-)^2}{4p_+p_-}\right)\\
\varepsilon_g\le \varepsilon\le -\frac{1}{2}&:\quad \nu'(\varepsilon) = 
\frac{4}{\pi \hbar^2 \omega\omega_0}
\left\{  \begin{array}{cc} 
\frac{2}{\sqrt{(y_+-z_-)(z_+-y_-)}} K\left(\frac{(y_+-z_+)(z_--y_-)}{(y_+-z_-)(z_+-y_-)}\right) , & g^2\mu\ge 1   \\    
 \frac{1}{\sqrt{(y_+-z_+)(y_--z_-)}} K\left(\frac{(y_+-y_-)(z_+-z_-)}{(y_+-z_+)(y_--z_-)}\right) , & g^2\mu\le 1   
\end{array} \right.\\
-\frac{1}{2}\le \varepsilon\le  \frac{1}{2}&:\quad \nu'(\varepsilon) = 
\frac{4}{\pi \hbar^2 \omega\omega_0}
\frac{1}{\sqrt{(y_+-y_-)(z_+-z_-)}} K\left(\frac{(y_+-z_+)(y_--z_-)}{(y_+-y_-)(z_+-z_-)}\right) . \label{denuexact_region2}
\end{align}
\end{subequations}
 \end{widetext}
Here, $y_\pm$ and $z_\pm$ have already been defined in \eq{yzdef}, 
\be
p_\pm \equiv \sqrt{(g^2\mu - y_\pm)^2 + 4g^2\mu (\varepsilon_g-\varepsilon)},
\ee
and $K(m)\equiv \int_0^{\pi/2}d\varphi (1-m\sin^2\varphi)^{-1/2}$ denotes the elliptic integral of the first kind.
Note that for the normal phase $\mu<1$ only \eq{denuexact_region2} is relevant.

\subsection{ESQPT for $0\le g\le 1$}
\begin{figure}[t]
\includegraphics[width=\columnwidth]{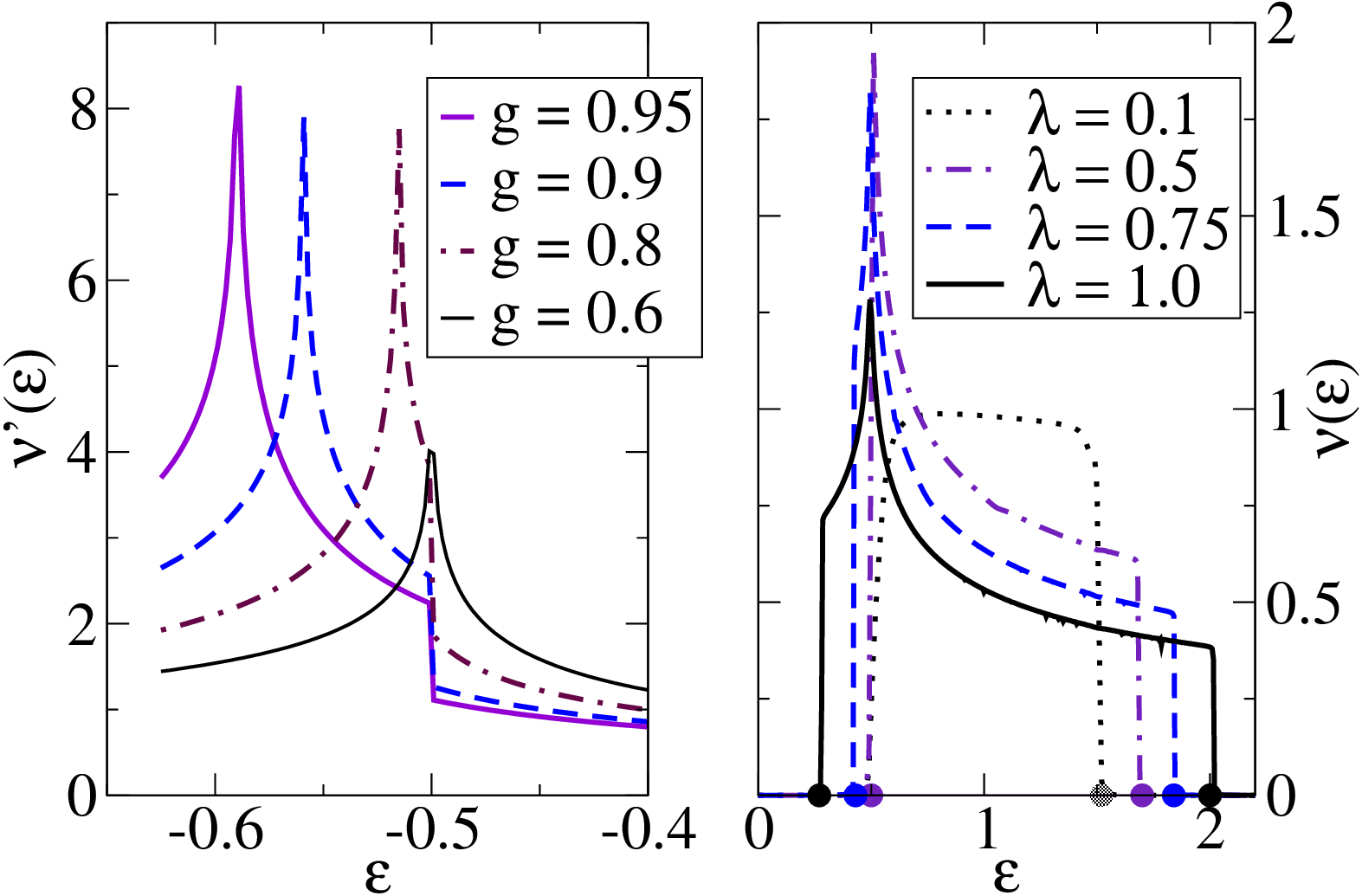}
\caption[]{\label{denuall}LEFT: DOS derivative $\nu'(\varepsilon)$, \eq{denuexact}, as a function of scaled energy $\varepsilon\equiv E/(N \hbar\omega_0)$ for generalized Dicke models  in the superradiant phase for various values of $g$, \eq{Hamiltonian} and criticality parameter $\mu=2$. RIGHT: DOS for the Tavis-Cummings model, \eq{nu_tavis_exact}, restricted to excitation number $N_{\rm ex}=N$, cf. \eq{Nexdef}, frequencies $\omega=2$, $\omega_0=1$ for various coupling parameters $\lambda$. The filled dots on the $\varepsilon$-axis indicate the values of the band-edges for the respective values of  $\lambda$ in the numerical data 
in Fig. 4 of reference \cite{PeresFernandezetal2011}. }
\end{figure} 
Figure \ref{denuall} (left) displays the main features contained in our expressions \eq{denuexact} in the superradiant phase. At small $g<1/\sqrt{\mu}$, only the log-type singularity appears at $\varepsilon=-\frac{1}{2}$, reflecting the ESQPT that we already saw in the Dicke ($g=0$) case and anticipated from the discussion of the potential \eq{potential}.
Writing the scaled energy $\varepsilon=-\frac{1}{2}+\delta$ with small $\delta$, from our exact expressions for $\nu'(\varepsilon)$  we explicitly extract (Appendix B) the logarithmic divergence
\be\label{logdiv}
\nu'\left(-\frac{1}{2}+\delta\right) \approx \frac{ - \ln \left|\frac {r}{16}   \delta\right|      }{\pi \hbar^2 \omega\omega_0 \sqrt{(\mu-1)(1-g^2\mu)}}
\ee
with the constant $r$ definined in \eq{rdef}.

This situation changes for larger $g>1/\sqrt{\mu}$, where the singularity at $\varepsilon=-\frac{1}{2}$ becomes a jump type discontinuity with 
a jump by a factor of $2$,
\be\label{denujump}
\nu'\left(\varepsilon =-\frac{1}{2}\pm 0^+\right) = \frac{3 \mp 1}{2 \hbar^2 \omega \omega_0\sqrt{(\mu-1) (g^2\mu-1)}}.
\ee
In addition, the log-type ESQPT singularity has now moved to the position $\varepsilon=\varepsilon_g$ corresponding to the two new saddle points in the potential landscape $U_+(x,p)$. 

In the limit $g=1$, from \eq{denuexact} we recover the Tavis-Cummings result \eq{nuTavis}: when pushed against the lower band-edge $\varepsilon_0$, all that remains from the log-singularity is a square-root divergence which (as discussed above) can be traced back to the Goldstone mode of the rotating-wave-approximation model in the superradiant phase.  
Another check is the ultrastrong Dicke ($g=0$) limit \eq{nuultrastrong} that follows from  \eq{denuexact}  for $\omega_0\to 0$.

\subsection{Collective excitations and degeneracies}\label{section_degeneracies}
In the normal phase $\mu<1$ and again in analogy with the Dicke case, we confirm the low-energy behavior 
\be\label{normalconfirm}
\nu'\left(\varepsilon =-\frac{1}{2}\right) = \frac{1}{\epsilon_+\epsilon_-}= \frac{1}{ \hbar^2 \omega \omega_0\sqrt{(\mu-1) (g^2\mu-1)}}
\ee
at arbitrary $0\le g\le 1$ with the collective low-energy excitation energies $\epsilon_\pm$. 
We checked that \eq{normalconfirm} also follows from the diagonalization of our Hamiltonian $\mathcal{H}$, \eq{Hamiltonian}, via a Bogoliubov-transformation (Appendix D),  or alternatively as the determinant of the Jacobian belonging to the normal phase fixed point in the classical equation-of-motion method \cite{BMSK2012}. 

In the superradiant phase $\mu>1$, we directly find from \eq{denuexact}, using $y_\pm = \mu$ at $\varepsilon=\varepsilon_0$, that at the lower band-edge
\be\label{lowerconfirm}
\nu'(\varepsilon_0)=  \frac{2}{\epsilon_+\epsilon_-} = \frac{2}{ \hbar^2 \omega\omega_0\sqrt{(1-g^2)(\mu^2-1)}}.
\ee
Again, we recover the  divergence of $\epsilon_+\epsilon_-$ at $g=1$ (Tavis-Cummings model) and our Dicke result for $g=0$, \eq{nulowDicke}. 
 
At the upper band-edge $\varepsilon=+\frac{1}{2}$, on the other hand, we have $y_-=z_-=-1$, $y_+=1+2\mu$ and $z_+=1+2g^2\mu$ and thus from \eq{denuexact}
\be\label{upperconfirm}
\nu'\left(\frac{1}{2}\right)=   \frac{1}{ \hbar^2 \omega\omega_0\sqrt{(1+\mu)(1+g^2\mu)}}.
\ee
Both forms \eq{normalconfirm}, \eq{lowerconfirm} are consistent with  the general form of the low-energy behavior of the model described by two collective modes with energies $\epsilon_\pm$, i.e., 
$\nu'(\varepsilon) \to g_{\rm d}/(\epsilon_+\epsilon_-)$ at the lower band-edge both in the normal and in the superradiant phase, where $g_{\rm d}$ is the level degeneracy factor, cf. \eq{lowenergy}.
Note that the classical potential \eq{potential} has $g_{\rm d}=2$ equivalent  minima at $\mu>1$ which is, of course, consistent with the {\em two} effective Hamiltonians describing the superradiant phase at low energies \cite{EB03two}. Here, the Tavis-Cummings case ($g=1$) can be formally interpreted as having degeneracy $g_{\rm d}=\infty$.

At this point, an interesting comparison can be made with recent numerical results by Puebla, Rela\~{n}o, and  Retamosa \cite{PRR13}, who found that in the superradiant phase, 
the energy levels in the Dicke model ($g=0$) are doubly degenerate ($g_{\rm d}=2$)  below the ESQPT critical energy $\varepsilon=-\frac{1}{2}$ and non-degenerate ($g_{\rm d}=1$) above that energy. In the normal phase, in contrast, they found no degeneracy at any energy. 

Our results above only refer to energies at the band edges, but they are consistent with this picture and generalize it to models with $g\ge 0$. In particular, the upper band-edge value \eq{upperconfirm} holds for all values of the criticality parameter $\mu$, in agreement with the absence of degeneracy at large energies found in \cite{PRR13}.

\subsection{Comparision with the LMG model}
As mentioned in the introduction, our results bear close analogies with the extensive studies of  Ribeiro, Vidal, and Mosseri \cite{RVM07,RVM08} for the Lipkin-Meshkov-Glick (LMG) model,
\be\label{Lipkin}
\mathcal{H} = -\frac{1}{N}\left(\gamma_x J_x^2 + \gamma_y J_y^2\right) - h J_z,
\ee
where in the $\gamma_y-\gamma_x$ phase diagrams four different phases were identified. 
Non-analyticities in the DOS $\nu(E)$ and the integrated DOS were  related to extremal points in the classical potential landscape belonging to \eq{Lipkin}, cf. our analysis in section \ref{potential_landscape}, and analytical expressions in terms of elliptic integrals followed via a mapping to a first-order non-linear differential equation.

For the Dicke-type models \eq{Hamiltonian}, due to the additional boson degree of freedom, the derivative $\nu'(E)$ of the level density (rather then $\nu(E)$ itself) is the key quantity in the analysis, but otherwise we have a clear correspondence: first, in the  normal (symmetric) phase both models have smooth level densities. Next, the single-log-divergence phase of the LMG model (phase II in \cite{RVM08} with $|\gamma_y| < h < \gamma_x$) corresponds to the case $g^2\mu\le 1$ in the Dicke models, whereas
the single-log/ jump phase of the LMG model (phase IV in \cite{RVM08} with $h < \gamma_y < \gamma_x$)  corresponds to the case $g^2\mu\ge 1$ in the Dicke-type models. In this latter phase, we 
obtain the same factor of two jump-discontinuity as \cite{RVM08}, cf. \eq{denuexact}, but with our method we can not further analyse the spectral subtleties there since we have no access to, e.g., the energy difference between two consecutive levels. Also note that we have only considered positive  couplings in  \eq{Hamiltonian} which is why there is no analogon to the phase III \cite{RVM08} with two log divergences in the LMG. 

Finally, the isotropic LMG model ($\gamma_y=\gamma_x$) is easily solvable in term of $J_z$ eigenstates only, and it has a single Goldstone mode \cite{VDB07}. In this limit, the LMG model corresponds to the (unrestricted) Tavis-Cummings Hamiltonian ($g=1$), cf. \eq{nuTavis} and section \ref{potential_landscape}.

\section{Restricted Tavis-Cummings model}\label{TCrestricted}
Finally, we turn to the Tavis-Cummings model ($g=1$) including the restriction defined by a fixed value of the conserved excitation number $N_{\rm ex}$, \eq{Nexdef}. P\'{e}rez-Fern\'{a}ndez and co-workers found a ground state QPT determined by the condition $\lambda >|\omega_0-\omega|/2$, and an ESQPT in the form of a strongly increased level density at $\varepsilon\equiv E/(N\hbar \omega_0)=+\frac{1}{2}$ and a needle-like singularity of the observable $\langle J_z \rangle$ at that energy  \cite{PeresFernandezetal2011}. Unfortunately and somewhat ironically, in contrast to a numerical analysis, the additional conserved quantity $N_{\rm ex}$ in the restricted Tavis-Cummings case $g=1$ makes it much harder to make analytical progress (when compared to all other models for $0\le g \le 1$ including the Dicke case $g=0$).

In our method based on the partition sum,  $\mathcal{Z}(\beta)$ now  has to be carried out at fixed $N_{\rm ex}$, a condition that can be included in the angular momentum trace part, \eq{partition}, in the form of a delta function reflecting \eq{Nexdef}, cf. Appendix A, leading to
\be\label{partitionTavis}
\mathcal{Z}(\beta) &=& \sum_{mm'} e^{-\beta \hbar m'\sqrt{\omega_0^2 + \frac{4\lambda^2}{N} \left( K-m\right)   }}\nonumber\\
&\times& e^{-\beta  \hbar \omega \left( K-m\right)   }\left|d_{mm'}(\theta)\right|^2 ,
\ee
where $K\equiv N_{\rm ex}-N/2$,
\be\label{ddef}
d_{mm'}(\theta) \equiv \langle m | e^{-i\theta J_y}| m'\rangle
\ee
is a rotation matrix element (Wigner's $d$-function) \cite{Brink_Satchler}, and the angle $\theta$ is defined by 
\be
\cos \theta = \frac{1}{\sqrt{1+ \frac{4\lambda^2}{N\omega_0^2}(K-m)}}.
\ee
As we are interested in the $N\equiv 2j \to \infty$ limit only, we  use the semiclassical approximation for the rotation matrix element \cite{Brink_Satchler,footnote_Wignerd}, 
\be\label{Wignerdsquare}
|d_{mm'}(\theta)|^2  \approx \frac{1}{\pi}\left[ j^2 \sin^2\theta - m^2 -m'^2 + 2mm' \cos \theta\right]^{-\frac{1}{2}},
\ee
which holds for positive arguments of the square-root, and where $|d|^2$ is approximated by zero otherwise. 

After converting the $m$- sums into integrals using $m\equiv Nx$, $m'\equiv Nx'$, this leads to
\be\label{nu_tavis_exact}
\nu(\varepsilon) &=& N\int_{-\frac{1}{2}}^{\frac{1}{2}}{dx} \frac{ \theta\left(\frac{1}{2} -|x'|\right)}{  \hbar\sqrt{ \omega_0^2 + 4\lambda^2 \left(    \frac{K}{N}-x \right)}}
|d_{Nx,Nx'}(\theta)|^2\nonumber\\
x'&\equiv& \frac{\varepsilon/\hbar- \omega \left( \frac{K}{N}-x\right) }{\sqrt{ \omega_0^2 + 4\lambda^2 \left(    \frac{K}{N}-x \right)}}.
 \ee
Note that in contrast to the DOS in the unrestricted cases discussed above, $\nu(\epsilon)$ is of order $N^0=1$ and thus not proportional to $N$ (the factor $N$ cancels with an $1/N$ from the Wigner $d$-function at large $N$). This corresponds to the reduction of dimensionality of the model due to the additional conserved quantity $N_{\rm ex}$ and is best visualized in a lattice representation of our model
$\mathcal{H}$, \eq{Hamiltonian}, where each point of the lattice represents a basis state $|jm\rangle \otimes |n\rangle$ \cite{Bra05}. The RWA-version $g=1$, i.e. the Tavis-Cummings model, then 
decomposes into independent, parallel chains that can be labeled by $N_{\rm ex}$ and that become one-dimensional lines in the thermodynamical limit, whereas the full lattice is two-dimensional. 

Results for the DOS $\nu(\varepsilon)$  for the restricted Tavis-Cummings model with conserved quantity $K=N/2$ ($N_{\rm ex}=N$), $\omega_0=1$, and $\omega=2$  are shown in Figure \ref{denuall}. For small $\lambda$, $\nu(\varepsilon)$ essentially has the shape of the uncoupled case where
\be
\nu(\varepsilon) = \frac{\theta\left(\frac{1}{2}-\left|\frac{\varepsilon/\hbar-\omega/2}{\omega_0-\omega}\right|\right)   }{\hbar \left|\omega_0-\omega\right|},
\ee   
which follows from the partition sum \eq{partitionTavis} for $\lambda=0$ and $d_{mm'}(\theta)=\delta_{mm'}$. 
At finite $\lambda$, we did not find a simple analytical form for the band-edges of $\nu(\varepsilon)$, but their numerical values following from \eq{nu_tavis_exact} agree well
with the results from exact numerical diagonalizations by P\'{e}rez-Fern\'{a}ndez {\em et al.} \cite{PeresFernandezetal2011}. 

In a similar way, we find a logarithmic singularity in $\nu(\varepsilon)$ at $\varepsilon=\frac{1}{2}$ for $\lambda>\lambda_c$, in agreement with the needlelike 
singularity  of $\langle J_z \rangle$ found in  \cite{PeresFernandezetal2011,footnote_Tavis}:
expanding the argument of the $d$-function \eq{Wignerdsquare} below the upper integration limit $x=\frac{1}{2}$, we find a purely quadratic behaviour
\be
\frac{\sin^2\theta}{4} - x^2 -x'^2 + 2xx' \cos \theta \approx \left(-1+4\lambda^2\right) \left(x-\frac{1}{2}\right)^2 
\ee
with no constant or linear term at $\varepsilon=\frac{1}{2}$, and thus $|d_{Nx,Nx'}(\theta)|^2 \propto | x-\frac{1}{2}   |^{-1}$ which upon integration leads to the logarithmic form for $\lambda>0.5$ there.

\section{Conclusion}
In all of our calculations, we have only considered the semiclassical limit  for which the partition sum $\mathcal{Z}(\beta)$ and thereby the DOS can be obtained without further approximations. 
Importantly, in order to arrive at our results we had to keep the full angular momentum character of the model, i.e., we 
did not make any kind of expansion using Holstein-Primakov bosons \cite{EB03two,Bra05}. 
The close analogy to results for the LMG model \cite{RVM08} suggests an equivalence between LMG and Dicke models not only for canonical thermodynamics \cite{TBZ09} but also for the `abnormal' microcanonical situation relevant for ESQPTs \cite{footnote_equivalence}.

For finite $N$, an obvious  next task would be to extract finite-size scaling exponents \cite{VD06,Wiletal12,PR13} for the ESQPT (cf. recent numerical results for the Dicke $g=0$ case \cite{PRR13}) 
or  an $1/N$-expansion similar to the LMG \cite{RVM07,RVM08} model. 

The Hellmann-Feynman theorem \eq{Jz} links the ESQPT non-analyticities to observables (or in fact the QPT order parameters), which might be more relevant for possible experiments than the DOS itself. 
Here, our analysis has remained incomplete in that we have only focused on the Dicke ($g=0$) case. In the $g>0$ case, the analytical evaluation of $\langle J_z\rangle $ is in principle straightforward, but 
for a comparison with the regime in which spectral subtleties similar to the LMG model \cite{RVM08} are expected one  would have to do quite some numerical efforts in addition. Another open point is 
the calculation of angular momentum observables (like $J_x$) that can not  directly be obtained via the  Hellmann-Feynman theorem.

An essential condition for the ESQPT in the Dicke models is the restriction to the Dicke states  $|jm\rangle$ with maximum $j=N/2$ in order to avoid the usual high degeneracy, i.e. the entropy term in $\mathcal{Z}(\beta)$ that leads to completely different physics, i.e. a thermal phase transition. In the ultrastrong coupling limit $\lambda\to \infty$ of the Dicke ($g=0$) model, we have recently discussed \cite{ABEB12} a realization of such a restriction with bosons, where the partition sum does not contain the combinatorial degeneracy factor of the usual fermionic (spin one-half) Dicke case and as a result, the thermal phase transition does not occur. An interesting option therefore would be to use bosons and  to directly explore the   properties  related to the thermodynamical ensemble defined by our canonicial  partition sum, \eq{partition}. In principle, one could then try to directly reconstruct ESQPT properties from equilibrium quantities at finite temperatures. 

A further point is the peculiar character of the models where ESQPTs have been studied so far. The Dicke or LMG models (which correspond to zero-dimensional field theories), are special in that there is no intrinsic length scale (like in lattice spin models). In the thermodynamic limit,  mean-field theory becomes exact and the ground state QPTs always follow some (classical) bifurcation scenario, on top of which one has non-trivial  finite-size corrections. A next step would therefore be to investigate ESQPTs in generic many-body systems in an expansion around a mean-field approximation (cf. \cite{KCU10} for a recent study of metastable QPT in a one-dimensional Bose gas).  

We also emphasize that the DOS $\nu(E)$ relevant for ESQPTs is different from the usual single quasiparticle excitation density of states known from, e.g., optical excitation spectra in many-body systems (cf. \cite{Zal12} for a recent example in the Bose-Hubbard Hamiltonian). Nevertheless, it would be worth to investigate the relation between the two quantities (be it only on a technical level) for further models in detail, in particular in view of the `band-structure' character of our calculation above. 

Finally a comment on possible experimental realizations of ESQPTs in Dicke models. Quantum quenches \cite{PeresFernandezetal2011,PRR13} seem to be a promising possibility to convert the singular energetic features into the time domain. The ground state QPT has been experimentally  tested both for the Dicke-Hepp-Lieb \cite{Dicke_experiment} and the LMG model \cite{Oberthaler_2010} in Bose-Einstein condensates. 
One challenge, as mentioned above, is to stay within the relevant sub-spaces of states (e.g. the Dicke states with $j=N/2$) when implementing the effective Hamiltonian $\mathcal{H}$, \eq{Hamiltonian}, for a `real' physical system.

\section*{Acknowledgements}
I thank P. P\'{e}rez-Fern\'{a}ndez for discussions on ESQPTs, for providing the original numerical data from Fig. 2 in \cite{PeresFernandezetal2011Jz} for the inversion $\langle J_z\rangle $, and 
for showing me his unpublished data from new numerical calculations for the photon number $\langle a^\dagger a\rangle$ in the Dicke model. 
I am also indebted to C. Emary, A. Rela\~no and P. Ribeiro for valuable comments on this manuscript, and I acknowledge support by the  DFG  via projects BR 1528/8-1 and SFB 910.

\begin{appendix}
\section{Angular momentum trace}
We evaluate the angular momentum trace $Z(\alpha;\beta))\equiv{{\rm Tr}}e^{-\beta H_g}$ in \eq{partition} by writing $\alpha=x+ip$ and
\be\label{Hg}
H_g \equiv \hbar\omega_0J_z+ \frac{2\hbar\lambda}{\sqrt{N}} \left ( x J_x - g p J_y \right).
\ee 
We carry out the trace by unitarily rotating the  angular momentum, first rotating around the $J_z$-axis via
\be
\gamma_1 (- J_y \sin \phi + J_x \cos \phi)  &=& e^{i\phi J_z} \gamma_1 J_x  e^{-i\phi J_z} 
\ee
with parameters $\gamma_1 \sin \phi =g p\frac{2\lambda}{\sqrt{N}}$, $\gamma_1 \cos \phi = x\frac{2\lambda}{\sqrt{N}}$ after which we rotate  the resulting  $\omega_0 J_z + \gamma_1J_x$ around $J_y$, using 
\be
\gamma (-J_x \sin \theta + J_z \cos \theta) &=& e^{i\theta J_y} \gamma J_z  e^{-i\theta J_y} 
\ee
and identifying $\omega_0 = \gamma \cos \theta$ and $ -\gamma \sin \theta = \gamma_1$. 
The resulting exponent  in the trace is now diagonal,
\be\label{trace_sum}
Z(\alpha;\beta)&=&{{\rm Tr}}e^{-\beta \hbar\gamma J_z}= \sum_{m=-N/2}^{N/2}  e^{-\beta m \gamma }
\ee
with the frequency $\gamma=\sqrt{ \gamma_1^2 + \omega_0^2}$ given by
\be\label{gammadef}
\gamma = \omega_0\sqrt{  1 +    \frac{4\lambda^2}{N\omega_0^2}(x^2+g^2p^2)        }.
\ee
Note that we can either use the positive or negative square-root for $\gamma$ as the sum is symmetric in  $m$. 
For large $N\to \infty$, we can neglect the difference between $N$ and $N+1$ to write
$Z(\alpha;\beta)=\sum_\pm \pm e^{\pm \beta \gamma \hbar N/2}/(e^{\beta\hbar \gamma}-1)$. In the semiclassical limit $\beta\hbar \omega_0\to 0$ this becomes
$\sum_\pm \pm e^{\pm \beta \gamma \hbar N/2}/(\beta \hbar \gamma) $, a result that one also obtains by replacing the 
sum \eq{trace_sum} by the integral $N \int_{-\frac{1}{2}}^{\frac{1}{2}} dm e^{-\beta N m \hbar\gamma} $.

Re-scaling of the integration variables $\tilde{x}\equiv x/\sqrt{N}$, $\tilde{p}\equiv g p/\sqrt{N}$, introducing polars and defining
$y\equiv \sqrt{1+ \frac{4\lambda^2}{\omega_0^2}r^2} $ now leads to \eq{partition2}.

For the Tavis-Cummings model ($g=1$) discussed in section \ref{TCrestricted}, the partition sum is restricted by a conserved excitation number $N_{\rm ex}$ which is included in the angular momentum trace in the form of a delta function,
\be
Z_{\rm TC}&\equiv&{{\rm Tr}}\left[\delta\left( K-|\alpha|^2-J_z\right)
 e^{-\beta H_g}\right],
\ee 
with the same $H_g$, \eq{Hg}, and $K\equiv N_{\rm ex}-N/2$.
Again, we first rotate the exponential by an angle $\phi$ around  the $J_z$ axis as above, but the second rotation by the angle $\theta$ around the $J_y$ axis does not commute with $J_z$ in the delta function, and therefore
\be
Z_{\rm TC}&=&{{\rm Tr}}\left[\delta\left( K-|\alpha|^2-J_z\right) 
e^{i\theta J_y} e^{- \beta\hbar \gamma J_z} e^{-i\theta J_y} \right]
\ee
with $\gamma$ given by  \eq{gammadef} for $g=1$. Here, the angle $\theta$ is given by
\be
\cos \theta = \frac{1}{\sqrt{1+ \frac{4\lambda^2}{N\omega_0^2} |\alpha|^2}}
\ee
The trace can be done explicitely by inserting a complete set of Dicke states $|jm\rangle$, leading to 
\be
Z_{\rm TC}&=&\sum_{mm'} \delta\left(K-|\alpha|^2-m\right)
\left|d_{mm'}(\theta)\right|^2  e^{- \beta \hbar \gamma m'},
\ee
where the matrix element $d_{mm'}$  is  Wigner's $d$-function \eq{ddef}. Inserting into \eq{partition} and carrying out the $\alpha$-integral then yields \eq{partitionTavis}.

\section{Dicke model}
The DOS in the Dicke case ($g=0$) follows from the Laplace back-transformation of the partition sum \eq{partitionDicke} by use of 
\be
\mathcal{L}^{-1}[\beta^{-\frac{3}{2}}e^{-\beta\Omega}](E) = \frac{2}{\sqrt{\pi}} \sqrt{E-\Omega} \theta(E-\Omega)
\ee
and writing $\Re \sqrt{x+i0} = \sqrt{x}\theta(x)$ (which is convenient for some of the following transformations),
\be
\nu(\varepsilon) = \frac{2N}{\pi  \sqrt{\mu}\omega } \Re \sum_\pm \pm \int_1^\infty dy \frac{ \sqrt{\varepsilon+i0 + \Phi_\pm(\frac{1}{\mu},y)} }{\sqrt{y^2-1}}.
\ee
Within the band, the explicit evaluation of the step function leads to  \eq{nuDickeintegral}.

Next and again within the band,   the mean inversion follows from \eq{Jz} as
\be
\langle J_z \rangle = -\frac{N^2}{2\pi \hbar \omega\mu\nu(\varepsilon)} I_-^{\frac{1}{2}},
\ee
where we defined the integrals (that we numerically evaluate to obtain the curves in Fig. \ref{Figobservables}), 
\be
 I_{\sigma}^\alpha&\equiv & \int_{y_0}^{y_+}
\frac{\sigma\frac{y^2-1}{\mu} +{y}}{\sqrt{y^2-1}}\left[(y_--y)(y-y_+)  \right]^{\alpha}dy
\ee
with the sign $\sigma=\pm $, $\alpha=\pm\frac{1}{2}$, and  the lower limit 
\be
y_0= y_-,\quad \varepsilon\le -\frac{1}{2};\quad y_0= 1,\quad \varepsilon\ge -\frac{1}{2}
\ee
and $y_\pm \equiv \mu \pm 2\sqrt{\mu}\sqrt{\varepsilon-\varepsilon_0}$. At $ \varepsilon = \frac{1}{2}$ we find $I_-^{\frac{1}{2}}=0$ and thus $\langle J_z \rangle =0$,  cf. \eq{Jzzero}. 

In the vicinity of the ESQPT, we write $\varepsilon=-\frac{1}{2}+\delta$. For $\delta\to 0$, we find
$\frac{\partial }{\partial \omega_0} \nu(\varepsilon) \approx \frac{1}{4}\nu'(\varepsilon)$ with  the logarithmic singularity \eq{nuDickelog} (also cf. \eq{logdiv}), and  as a consequence
the derivative of $\langle J_z\rangle(\varepsilon)$ is given by
\be\label{Jzderivative}
\frac{\partial}{\partial \varepsilon} \langle J_z\rangle(\varepsilon) &=& - \frac{\frac{\partial }{\partial \varepsilon} \nu(\varepsilon) \langle J_z\rangle(\varepsilon) + \frac{\partial }{\partial \omega_0} \nu(\varepsilon)     }{\nu(\varepsilon)}\nonumber\\
& \propto&    
\log|\delta|,
\ee
with $\frac{\partial }{\partial \omega_0} \nu(E) =\frac{N}{2\pi\omega\omega_0} I_-^{-\frac{1}{2}} $.
As we checked numerically, the logarithmic singularity near $\varepsilon=-\frac{1}{2}$ in \eq{Jzderivative} has a prefactor that (depending on the value of $\mu$) is either positive or negative. 

For the boson number $\hat{n}$ , we used the Hellmann-Feynman theorem  to find the first moment
\be
\langle \hat{n} \rangle(\varepsilon) = \frac{N \omega_0 }{\omega} \left( \frac{\varepsilon}{3}+ \frac{N}{6\pi \mu \hbar \omega  \nu( \varepsilon)}I_+^{\frac{1}{2}}\right),\quad  \varepsilon\le \frac{1}{2}
\ee
and the linear form  \cite{AH_NJP12} $\langle \hat{n}\rangle (\varepsilon) =  N   \frac{\omega_0}{\omega} \left( \frac{\mu}{6}+ \varepsilon\right)$ for $\varepsilon\ge \frac{1}{2}$, where the constant follows from $I_+^{\frac{1}{2}}=\pi \mu (2+\mu)$ at $\varepsilon = \frac{1}{2}$.
We obtain the second moment  via \eq{QLaplace} by Laplace-backtransformation and carrying out the integration;
\be
\langle \hat{n}^2 \rangle(\varepsilon) = \left(\frac{ N\omega_0}{\omega}\right)^2    \frac{N}{\pi \hbar\omega \nu(\varepsilon)} \left[ \frac{J_2^{\frac{1}{2}} }{\mu}  +  \frac{J_0^{\frac{5}{2}}}{80\mu^3}
+  \frac{J_1^{\frac{3}{2}}}{6\mu^2}        \right],
\ee
where we defined the integrals (to be solved numerically)
\be
J_{\sigma}^{\alpha} \equiv\int_1^\infty dy \frac{ \Re \sum_\pm \pm \left[ (y_1^\pm-y)(y-y_2^\pm)+i0^+\right]^\alpha }{(4\mu)^\sigma( y^2-1)^{\frac{1}{2} -\sigma}},
\ee
with $y_{1,2}^\sigma \equiv  {\sigma \mu} \mp \sqrt{ \mu^2  +1 + {4\varepsilon \mu } }$, $\sigma=\pm$ where the index $1$ ($2$) belongs to the negative (positive) root.

\section{Logarithmic singularities in  $\nu'(\varepsilon)$ }
In the superradiant regime ($\mu>1$) we first consider $g^2\mu<1$ near the ESQPT, writing the scaled energy 
$\varepsilon=-\frac{1}{2}+\delta$ with small $\delta$. Expanding $y\pm$ and $z_\pm$, \eq{yzdef}, in $\delta$, one finds for the arguments of the elliptic integral in \eq{denuexact}
\be
\frac{(y_+-z_+)(y_--z_-)}{(y_+-y_-)(z_+-z_-)} &=& 1 + r  \delta + O(\delta^2),\quad \delta>0\\
\frac{(y_+-y_-)(z_+-z_-)}{(y_+-z_+)(y_--z_-)} &=& 1 - r  \delta + O(\delta^2),\quad \delta<0
\ee
with the parameter
\be\label{rdef}
r\equiv \frac{2\mu^2-g^2\mu(1+\mu)}{2(\mu-1)^2(g^2\mu-1)},
\ee
and we use $K(1-|x|) = -\frac{1}{2} \ln \frac{|x|}{16} + O(x)$ to arrive at \eq{logdiv}. 

At $g^2\mu>1$, the log-singularity moves to the energy $\varepsilon=\varepsilon_g$, \eq{epsgdef}, where
$z_\pm = g^2\mu$ and thus 
\be
\frac{(y_+-z_+)(z_--y_-)}{(y_+-z_-)(z_+-y_-)}\to 1.
\ee
The value unity in the argument of the elliptic integral $K$ again denotes the appearance of a logarithmic singularity. Finally, we numerically checked that the argument
of $K$ 
\be
\frac{(y_+-y_-)^2-(p_+-p_-)^2}{4p_+p_-} =\theta(g^2\mu-1),\quad \varepsilon=\varepsilon_g,
\ee
which confirms that also for energies just below $\varepsilon_g$, we have a logarithmic divergence for $g^2\mu>1$.

\section{Bogoliubov transformation (normal phase)}
To extract the collective excitation energies $\epsilon_\pm$ in the normal phase, we use the Holstein-Primakoff representation with a bosonic mode created by $b^\dagger$ \cite{EB03two,Bra05},
\be
J_+= b^\dagger \sqrt{N-b^\dagger b},\quad J_z = b^\dagger b -\frac{N}{2}
\ee 
and expand the Hamiltonian \eq{Hamiltonian} for large $N$ which leads us to 
\be
\mathcal{H}&=&\hbar \omega a^\dagger a + \hbar  \omega_0 b^\dagger b+   \hbar \left( \lambda_+  a  b^\dagger  +  \lambda_-  a  b  +  H. c. \right),
\ee
where we defined $\lambda_\pm \equiv  \lambda \frac{1 \pm g}{2}$ and omitted a constant.
We write $\mathcal{H}$ in  canonical form \cite{HY00}
$
\mathcal{H}/\hbar= \mathbf{a}^\dagger \alpha \mathbf{a} + \frac{1}{2} \mathbf{a}^\dagger \gamma \tilde{\mathbf{a}}^\dagger
+ \frac{1}{2} \tilde{\mathbf{a}}\gamma^\dagger \mathbf{a}
=  \frac{1}{2} \Lambda N \Sigma \tilde{\Lambda} - \frac{1}{2}{\rm Tr} \alpha,
$ 
with vectors $\mathbf{a}^\dagger\equiv(a^\dagger,b^\dagger)$, $\tilde{\mathbf{a}}\equiv (a,b)$, 
$\Lambda \equiv (\mathbf{a}^\dagger,\tilde{\mathbf{a}})$, 
and where the $\tilde{.}$ converts rows into columns and vice versa. 
Here, we used the canonical commutation relations, written in dyadic form with the $4\times 4$ 
symplectic unity $\Sigma$ as
\be\label{canonical}
[ \tilde{\Lambda},\Lambda] = \Sigma^{-1}= \twomatrix{0}{-1}{1}{0},
\ee
and we defined the matrices
\be
N\equiv \twomatrix{\alpha}{-\gamma}{\gamma^\dagger}{-\alpha}, \alpha \equiv  \twomatrix{\omega}{\lambda_+}{\lambda_+}{\omega_0},
\gamma \equiv \twomatrix{0}{\lambda_-}{\lambda_-}{0}.
\ee
As in classical mechanics, a canonical transformation $\Lambda'=\Lambda M$ leaves \eq{canonical} invariant if $\tilde{M} \Sigma^{-1} M =\Sigma^{-1} $, i.e. if $M$ is symplectic.
The eigenvalues of $\mathcal{H}$ follow from the eigenvalues of $N$, which come in pairs $\pm \epsilon_\pm$. The product $\epsilon_+\epsilon_-$ thus simply follows from the determinant of $N$,
\be
\det N &=& (\epsilon_+\epsilon_-/\hbar^2)^2 = \lambda^4 (1-\mu^{-1})(g^2 - \mu^{-1}) \nonumber\\
&=& (\omega \omega_0)^2 (\mu -1) (g^2 \mu -1),
\ee 
where in the last step we used $\mu = \lambda^2/(\omega \omega_0)$, \eq{mudef}, which confirms \eq{normalconfirm}. 

\end{appendix}


\begin{thebibliography}{10}
\bibitem{footnote_coupling}
{Our choice of coupling $\lambda$ thus corresponds to half the coupling used in
  \cite{EB03two}.}
\bibitem{footnote_Laplace}
{By shifting integration variables, we have $\mathcal{Z}(\beta) = e^{-\beta
  E_0}\int_{0}^\infty dE e^{-\beta E} \tilde{\nu}(E) $ and thus $\nu(E) \equiv
  \tilde{\nu}(E-E_0) = \mathcal{L}^{-1}[\mathcal{Z}(\beta) e^{\beta E_0}
  e^{-\beta E_0} ](E) =\mathcal{L}^{-1}[\mathcal{Z}(\beta)](E)$, where $
  \mathcal{L}^{-1}[...]$ denotes the inverse Laplace transformation. Here, it
  is understood that $\nu(E)$ vanishes below $E_0$. One obtains the value of
  $E_0$ (which is known for $\mathcal{H}$, \eq{Hamiltonian}, anyway) from an
  analysis of $\mathcal{Z}(\beta)$ for $\beta \to \infty$ (not discussed
  here)}.
\bibitem{footnote_Dicke_states}
{Note that this has to be contrasted with the usual calculations for the Dicke
  phase transition at finite temperature $T$ \cite{WH73,CGW73} where the trace
  always runs over all the $2^N$ configurations of the pseudo-spin
  $\frac{1}{2}$ states. }.
\bibitem{footnote_thermo}
{\eq{Jzthermo} is an example of the relation $\left.\frac{\partial x}{\partial
  y}\right|_z \left.\frac{\partial y}{\partial z}\right|_x \left.\frac{\partial
  z}{\partial x}\right|_y =-1$ often used in thermodynamics, if we write it in
  the form $\left.\frac{\partial \nu}{\partial E}\right|_{\omega_0}
  \left.\frac{\partial E}{\partial \omega_0}\right|_\nu \left.\frac{\partial
  \omega_0}{\partial \nu}\right|_E =-1$ and take $E=E_0$, the ground state
  energy. }.
\bibitem{Jzfit}
{Fits with $\langle J_z\rangle(\varepsilon)/N= -0.12+0.12
  |\varepsilon+0.5|^{1.5}$ for $\varepsilon>-0.5$ and $\langle
  J_z\rangle(\varepsilon)/N= -0.12+0.05 |\varepsilon+0.5|^{0.25}$ for
  $\varepsilon<-0.5$ \cite{PeresFernandezetal2011Jz} with parameters tuned by
  `look of the eye' work quite well within the band.}
\bibitem{footnote_Tavis}
{We did not explicitly calculate $\langle J_z \rangle$ via the Hellmann-Feynman
  theorem \eq{Jz}, which is -- in contrast to the Dicke case -- cumbersome for
  the restricted Tavis-Cummings model}.
\bibitem{footnote_equivalence}
{It it not clear, though, what such an equivalence would mean in practical
  terms when calculating certain quantities in either of the two models}.
\end{thebibliography}

\end{document}